\documentclass[preprint,12pt]{elsarticle}

\usepackage{amssymb}
\usepackage{amsmath}

\journal{Acta Materialia}

\begin{document}

\begin{frontmatter}



\title{Dynamic recrystallization in adiabatic shear banding: effective-temperature model and comparison to experiments in ultrafine-grained titanium}


\author{Charles K. C. Lieou and Curt A. Bronkhorst}

\address{Theoretical Division, Los Alamos National Laboratory, Los Alamos, NM 87545, USA}

\begin{abstract}
Dynamic recrystallization (DRX) is often observed in conjunction with adiabatic shear banding (ASB) in polycrystalline materials. The recrystallized nanograins in the shear band have few dislocations compared to the material outside of the shear band. In this paper, we reformulate the recently-developed Langer-Bouchbinder-Lookman (LBL) continuum theory of polycrystalline plasticity and include the creation of grain boundaries. While the shear-banding instability emerges because thermal heating is faster than heat dissipation, recrystallization is interpreted as an entropic effect arising from the competition between dislocation creation and grain boundary formation. We show that our theory closely matches recent results in sheared ultrafine-grained titanium. The theory thus provides a thermodynamically consistent way to systematically describe the formation of shear bands and recrystallized grains therein.
\end{abstract}

\begin{keyword}
Constitutive behavior \sep Dynamic recrystallization \sep Dynamic recovery \sep Shear banding \sep Titanium


\end{keyword}

\end{frontmatter}


\section{Introduction}
\label{sec:1}

Adiabatic shear banding (ASB) is the physical process by which plastic deformation localizes to a narrow region of a polycrystalline material. A mechanical signature of ASB is the concurrent, dramatic drop in the load-carrying capacity of the material as a function of displacement or strain. Often accompanying ASB is dynamic recrystallization (DRX), the microstructural evolution inside the shear band whereby recrystallization, possibly under the influence of temperature and mechanical load, creates tiny, nano-sized grains that have few defects, if not dislocation-free~\cite{dodd_2012,meyers_2003,rittel_2008,osovski_2012,osovski_2013,li_2017}. Because the shear-banding instability weakens the material and often precedes the development of voids and cracks as well as material failure, it is crucial to understand the physical mechanisms behind ASB and develop useful, predictive constitutive descriptions of these phenomena for various applications.

There have been varied theoretical, mathematical, and numerical attempts to model ASB, many of which are summarized in \cite{wright_2002} and \cite{dodd_2012}. Much of the existing literature on ASB (e.g., \cite{hines_1997,mourad_2017}) relies on phenomenological fits between stresses and strains, which do not shed much light on the underlying physical and micromechanical mechanisms for microstructural evolution and the onset of shear localization, and are therefore inadequate for an in-depth understanding that we hope to pursue.

Recently, a thermodynamic dislocation theory \cite{langer_2010,langer_2015} has been proposed by Langer, Bouchbinder, and Lookman (LBL) to describe strain hardening in polycrystalline solids. The statistical theory invokes the concept of an effective temperature, which quantifies the structural disorder that is responsible for driven atomic rearrangements underlying dislocation motion. In the LBL theory, the creation of dislocations and their subsequent motion are simply driven processes that dynamically minimize the free energy, in accordance with the second law of thermodynamics. Preliminary extensions of this theory \cite{langer_2016,langer_2017a} demonstrate that ASB is simply a runaway instability originating from structural heterogeneities, and resulting from thermal softening, or the inability for heat to dissipate as quickly as it is generated by plastic deformation. In one of these papers \cite{langer_2016}, Langer alludes to the possibility of DRX as the main driver of softening and the ASB instability -- first pointed out by \cite{rittel_2008} -- but makes no attempt to incorporate such structural changes into the set of dynamical variables in the theory. In order to gain deeper insight into the causal relations between thermal softening, DRX, and ASB, it is necessary to include the structural features responsible for recrystallization -- namely, grain boundaries -- into our formulation as a state variable, in addition to the coarse-grained dislocation density. As such, DRX can be understood as an entropic effect, in the same manner as dislocation creation and strain hardening.

The rest of this paper is organized as follows. In Section \ref{sec:2} we provide a summary of the LBL theory of dislocations, setting the stage for Section \ref{sec:3}, where we include the grain boundary density as an additional state variable to facilitate a description of grain size evolution. After accounting for the interaction between grain boundaries and dislocation lines, we specialize in Section \ref{sec:4} to a simplified two-zone model amenable to easy numerical calculations. There, we explore the role of DRX, and show that recrystallized grains with a reduced dislocation density emerge naturally in the shear band. We conclude in Section \ref{sec:5} with important open questions and future directions.

\section{Effective-temperature theory of dislocations: summary of basic equations}
\label{sec:2}

In this section, we provide a summary of the LBL theory of dislocations. The development here largely follows \cite{langer_2010,langer_2015}; we however do not specialize to simple geometries, but instead use the tensor notation to describe more general modes of deformation.

Consider a slab of the polycrystalline material with area $A$ and thickness $L$ equal to the typical length of a dislocation line. Let $s_{ij}$ and $\dot{\gamma}_{ij}$ be the deviatoric stress and the total deviatoric strain rate, respectively. These are related to the total stress tensor $\sigma_{ij}$ and the total strain rate $\dot{\epsilon}_{ij}$ through the relations
\begin{equation}
 s_{ij} = \sigma_{ij} - \dfrac{1}{3} \sigma_{kk} \delta_{ij}, \quad \dot{\gamma}_{ij} = \dot{\epsilon}_{ij} - \dfrac{1}{3} \dot{\epsilon}_{kk} \delta_{ij} .
\end{equation}
Define the stress invariant
\begin{equation}
 \bar{s} = \sqrt{\dfrac{1}{2} s_{ij} s_{ij}} .
\end{equation}
Then the Orowan relation, which gives the plastic strain rate $\dot{\gamma}_{ij}^{\text{pl}}$ in terms of the average speed $v$ of dislocations, reads
\begin{equation}
 \dot{\gamma}_{ij}^{\text{pl}} = \dfrac{\rho}{2} \dfrac{s_{ij}}{\bar{s}} b v ,
\end{equation}
where $\rho$ is the dislocation density, and $b$ is the magnitude of the Burgers vector.

Now the dislocation speed $v$ is related to the average spacing $l = 1 / \sqrt{\rho}$ between dislocations and the depinning rate $1 / \tau_P$ through $v = l / \tau_P$. Assume that depinning is a thermally activated process, so that
\begin{equation}
 \dfrac{1}{\tau_P} = \dfrac{1}{\tau} \exp \left( - \dfrac{1}{\theta} e^{- \bar{s} / s_T} \right).
\end{equation}
Here, $\tau \sim 10^{-12}$ s is the time scale of atomic vibrations, $\theta = T / T_P$ is the dimensionless thermal temperature measured in units of the depinning energy $k_B T_P$ ($k_B$ being the Boltzmann constant), and $s_T$ is the depinning stress, which simply equals the Taylor stress:
\begin{equation}
 s_T = \mu_T b \sqrt{\rho} \equiv \mu_T \sqrt{\tilde{\rho}} ,
\end{equation}
which conveniently defines the dimensionless dislocation density $\tilde{\rho}$. Here $\mu_T$ is about 1/30 times the shear modulus. Combining everything together, we find
\begin{equation}
 \tau \dot{\gamma}_{ij}^{\text{pl}} = \dfrac{1}{2} \dfrac{s_{ij}}{\bar{s}} \sqrt{\tilde{\rho}} \exp \left[ - \dfrac{1}{\theta} e^{- \bar{s} / ( \mu_T \sqrt{\tilde{\rho}})} \right] .
\end{equation}
Writing $\bar{\dot{\gamma}}^{\text{pl}} \equiv \sqrt{\frac{1}{2} \dot{\gamma}_{ij}^{\text{pl}} \dot{\gamma}_{ij}^{\text{pl}}}$, a convenient measure of the dimensionless strain rate is
\begin{equation}\label{eq:q}
 q \equiv 2 \tau \bar{\dot{\gamma}}^{\text{pl}} = \sqrt{\tilde{\rho}} \exp \left[ - \dfrac{1}{\theta} e^{- \bar{s} / ( \mu_T \sqrt{\tilde{\rho}})} \right] .
\end{equation}
Then the ratio of the invariant stress $\bar{s}$ to the depinning stress $s_T$ is
\begin{equation}
 \dfrac{\bar{s}}{\mu_T \sqrt{\tilde{\rho}}} = \ln \left(\dfrac{1}{\theta} \right) - \ln \left[ \ln \left( \dfrac{\sqrt{\tilde{\rho}}}{q}\right) \right] \equiv \nu ,
\end{equation}
which will be useful in subsequent analysis.

To complete the theoretical description we need evolution equations for the stress, temperature, effective temperature, and dislocation density. The evolution equation for the stress is simply a statement of linear elasticity, assuming nontrivially that the total strain rate $\dot{\gamma}_{ij}$ is a sum of elastic and plastic parts:
\begin{equation}\label{eq:s}
 \dot{s}_{ij} = 2 \mu (\dot{\gamma}_{ij} - \dot{\gamma}_{ij}^{\text{pl}} ),
\end{equation}
where $\mu$ is the shear modulus.

The thermal temperature $\theta$ increases with the work of deformation $\sigma_{ij} \dot{\epsilon}^{\text{pl}}_{ij}$; there is also heat exchange with the surroundings at temperature $\theta_0$, as well as heat conduction within the material:
\begin{equation}\label{eq:theta}
 \dot{\theta} = K \sigma_{ij} \dot{\epsilon}_{ij}^{\text{pl}} + K_1 \nabla^2 \theta - K_2 (\theta - \theta_0) .
\end{equation}
Here, $K = \beta / (T_P c_v)$, where $\beta$ is the fraction of work converted into heat, often taken to be around $0.9$, and $c_v$ is the specific heat capacity per unit volume of the polycrystalline material. $K_1$ and $K_2$ are dimensionless constants. Because shear banding in a polycrystalline material is a runaway instability associated with heat concentration, equation \eqref{eq:theta} should play an important role.

The second law of thermodynamics mandates that in order to minimize the free energy, the dislocation density $\tilde{\rho}$ must approach some steady state controlled by the effective temperature $\chi$: specifically, $\tilde{\rho}^{\text{ss}} = e^{-e_D / \chi}$, where $e_D$ is the typical formation energy of a single dislocation line \cite{langer_2010,langer_2015}. The effective temperature $\chi$ is defined as the derivative of the configurational energy $U_C$ with respect to the configurational entropy $S_C$, to be discussed in greater detail in the next Section; $\chi$ itself approaches some nonequilibrium steady state $\chi_0$ at a rate proportional to the work of shear deformation. In the linear approximation, this can be written as
\begin{equation}
 c^{\text{eff}} \dot{\chi} = s_{ij} \dot{\gamma}_{ij}^{\text{pl}} \left( 1 - \dfrac{\chi}{\chi_0} \right),
\end{equation}
where $c^{\text{eff}}$ is an effective specific heat capacity. Defining the dimensionless effective temperature $\tilde{\chi} \equiv \chi / e_D$, this evolution equation becomes
\begin{equation}\label{eq:chi}
 \dot{\tilde{\chi}} = \dfrac{\kappa_2}{\mu_T} s_{ij} \dot{\gamma}_{ij}^{\text{pl}} \left(1 - \dfrac{\tilde{\chi}}{\tilde{\chi}_0} \right) ,
\end{equation}
where $\kappa_2$ is a dimensionless quantity.

Finally, the evolution equation for the dislocation density reads
\begin{equation}\label{eq:rho}
 \dot{\tilde{\rho}} = \dfrac{\kappa_1}{\nu^2} \dfrac{s_{ij} \dot{\gamma}_{ij}^{\text{pl}}}{\mu_T} \left( 1 - \dfrac{\tilde{\rho}}{\tilde{\rho}^{\text{ss}} (\tilde{\chi})} \right) .
\end{equation}
The factor $1 / \nu^2$ is inserted for a self-consistent description of strain hardening; see for example \cite{langer_2010} for details. $\kappa_1$, which determines the fraction of input work converted into dislocations, is a dimensionless quantity which may depend on the grain size as well as the strain rate. This will be discussed in greater detail in the next section.

\section{Dynamic recrystallization as an entropic effect}
\label{sec:3}

Dynamic recrystallization is frequently observed in conjunction with, and often preceding, adiabatic shear banding~\cite{wright_2002,dodd_2012,rittel_2008}. Specifically, the material within the shear band often recrystallizes, forming smaller grains, under plastic strain. To describe DRX, we need to include the density of grain boundaries $\xi$ along some line that cuts through the material, or equivalently the grain boundary area per unit volume, as an internal variable, along with entities above such as the dislocation areal density or length per unit volume $\rho$. The reciprocal of the grain boundary density $\xi$ gives the characteristic grain size $d$.

\subsection{Nonequilibrium thermodynamics}

To this end, both the configurational energy per unit volume $U_C$ and the configurational entropy per unit volume $S_C$ of the material ought to depend not only on the dislocation density $\rho$, but also the grain boundary density $\xi$. Thus, 
\begin{eqnarray}
 \label{eq:U_C} U_C (S_C, \rho, \xi) &=& U_0 (\rho) + U_G (\xi) + U_{\text{int}} (\rho, \xi) + U_1 (S_1) ; \\
 \label{eq:S_C} S_C (U_C, \rho, \xi) &=& S_0 (\rho) + S_G (\xi) + S_1 (U_1).
\end{eqnarray}
Here, $U_0$ and $S_0$ are respectively the dislocation energy and entropy per unit volume, $U_G$ and $S_G$ are respectively the grain boundary energy and entropy per unit volume, and $U_{\text{int}}$ is the interaction energy per unit volume between grain boundaries and dislocation lines. We assume that the contributions from dislocations and grain boundaries to the configurational entropy are independent of one another. $U_1$ and $S_1$ are the contributions to the configurational energy and entropy densities from degrees of freedom other than dislocations and polycrystallinity; such contributions could arise from twinning, for example, which we do not consider explicitly for the time being. The configurational degrees of freedom of the polycrystalline material is coupled to the kinetic-vibrational degrees of freedom, which serves as the thermal reservoir, with energy and entropy $U_R$ and $S_R$. Then the first law of thermodynamics says that the total energy density $U_{\text{tot}}$ changes according to the work of deformation; that is,
\begin{eqnarray}
 \dot{U}_{\text{tot}} &=& s_{ij} \dot{\gamma}_{ij}^{\text{pl}} = \dot{U}_C + \dot{U}_R \\ &=& \chi \dot{S}_C + \left( \dfrac{\partial U_C}{\partial \rho} \right)_{S_C} \dot{\rho} + \left( \dfrac{\partial U_C}{\partial \xi} \right)_{S_C} \dot{\xi} + \theta \dot{S}_R .
\end{eqnarray}
(Here, we have used the thermodynamic definition of the effective temperature: $\chi = (\partial U_C / \partial S_C)_{\rho,\xi}$.) The second law of thermodynamics says that the entropy must be a non-decreasing function of time, i.e.,
\begin{equation}
 \dot{S}_C + \dot{S}_R \geq 0.
\end{equation}
Multiplying this by the effective temperature $\chi$ and eliminating $\dot{S}_C$, we find that
\begin{equation}\label{eq:second_law}
 {\cal W} + (\chi - \theta) \dot{S}_R \geq 0 ,
\end{equation}
where the dissipation rate is
\begin{equation}\label{eq:W}
 {\cal W} \equiv s_{ij} \dot{\gamma}_{ij}^{\text{pl}} - \left( \dfrac{\partial U_C}{\partial \rho} \right)_{S_C} \dot{\rho} - \left( \dfrac{\partial U_C}{\partial \xi} \right)_{S_C} \dot{\xi} .
\end{equation}
Each independently variable term in equations \eqref{eq:second_law} and \eqref{eq:W} must be nonnegative, according to the Coleman-Noll argument \cite{coleman_1963}. Thus, from equation \eqref{eq:second_law} we find that $(\chi - \theta) \dot{S}_R \geq 0$, from which we infer that the heat flux between the polycrystalline material and the heat reservoir must satisfy $Q \equiv \theta \dot{S}_R \propto (\chi - \theta)$. The first term in the dissipation rate is proportional to $s_{ij} \dot{\gamma}_{ij}^{\text{pl}}$, which is generally nonnegative. Then we arrive at the constraints
\begin{equation}\label{eq:U_C_constraint}
 - \left( \dfrac{\partial U_C}{\partial \rho} \right)_{S_C} \dot{\rho} \geq 0; \quad - \left( \dfrac{\partial U_C}{\partial \xi} \right)_{S_C} \dot{\xi} \geq 0 .
\end{equation}
Because
\begin{eqnarray}
 \left( \dfrac{\partial U_C}{\partial \rho} \right)_{S_C} &=& \dfrac{\partial U_0}{\partial \rho} + \dfrac{\partial U_{\text{int}}}{\partial \rho} - \chi \dfrac{\partial S_0}{\partial \rho} \equiv \dfrac{\partial F_C}{\partial \rho} ; \\
 \left( \dfrac{\partial U_C}{\partial \xi} \right)_{S_C} &=& \dfrac{\partial U_G}{\partial \xi} + \dfrac{\partial U_{\text{int}}}{\partial \xi}- \chi \dfrac{\partial S_0}{\partial \xi} \equiv \dfrac{\partial F_C}{\partial \xi} ,
\end{eqnarray}
where the configurational free energy density $F_C$ is given by
\begin{equation}\label{eq:F_C}
 F_C (\rho, \xi) = U_0 (\rho) + U_G (\xi) + U_{\text{int}} (\rho, \xi) - \chi (S_0 (\rho) + S_G (\xi) ),
\end{equation}
the thermodynamic constraint given by equation \eqref{eq:U_C_constraint} simply says that the dislocation density $\rho$ and the grain boundary density $\xi$ evolve in such a way as to minimize the free energy, approach some stationary values $\rho^{\text{ss}}$ and $\xi^{\text{ss}}$ given by the free energy minima.

Let us proceed to calculate $\rho^{\text{ss}}$. We already know that in the noninteracting dislocation approximation, the energy density of dislocations is $U_0 = e_D \rho / b$, where $e_D$ is the typical formation energy of a dislocation line of length $b$ equal to the length of the Burgers vector defined earlier. Meanwhile, a simple counting of the number of microstates through the number of sites occupied by dislocations yields $S_0 = (1/b) [ - \rho \ln ( b^2 \rho) + \rho]$. Let us neglect the interaction $U_{\text{int}}(\rho, \xi)$ between dislocations and grain boundaries for the time being. Thus, upon minimizing the free energy we find the familiar result $\rho^{\text{ss}} = (1/b^2) e^{- e_D / \chi}$, or 
\begin{equation}\label{eq:rho_ss}
 \tilde{\rho}^{\text{ss}} = e^{- e_D / \chi}.
\end{equation}

Similarly, the energy density of grain boundaries is given by $U_G =  e_G \xi / b^2$, where $e_G$ is an energy scale. Counting the number of sites cut through by a grain boundary, meanwhile, gives $S_G = (1 / b^2) [- \xi \ln (b \xi) + \xi]$. Then the steady-state density of grain boundaries is simply $\xi^{\text{ss}} = (1/b) e^{- e_G / \chi}$, or ($\tilde{\xi} \equiv b \xi$):
\begin{equation}\label{eq:xi_ss}
 \tilde{\xi}^{\text{ss}} = e^{ - e_G / \chi}.
\end{equation}
Inverting this gives the steady-state grain size
\begin{equation}\label{eq:d_ss}
 d^{\text{ss}} = b \exp \left( \dfrac{e_G}{\chi} \right) .
\end{equation}
One can also arrive at this result by enumerating the number of possible microstates to pack $N$ atoms into $N_G$ grains. In a 1-D polycrystal, this is equivalent to the ``stars-and-bars'' combinatorial problem of placing $N_G -1$ dividers in between a line of $N - N_G$ stars.

To describe grain size evolution and dynamic recrystallization, we need an evolution equation for the grain boundary density $\xi$ as new grain boundaries are created, or equivalently an evolution equation for the grain size $d$. But first let us consider the implications of equations \eqref{eq:rho_ss} and \eqref{eq:xi_ss}.

\subsection{Interaction between dislocation lines and grain boundaries; correction to the dislocation density}

A number of authors (e.g., \cite{rittel_2008,meyers_2003,li_2017}) observe DRX in conjunction with adiabatic shear banding; the DRX grains in the shear band apparently contain many fewer dislocations than the non-recrystallized grains. This observation is in contrast to the prediction of equations \eqref{eq:rho_ss} and \eqref{eq:d_ss}, which imply that as deformation proceeds and raises the effective temperature $\chi$, grains progressively become smaller with increasing dislocation densities. The fact that recrystallized grains with few dislocations is the entropically preferred configuration suggests that the interaction energy density $U_{\text{int}}$ between dislocation lines and grain boundaries must be nonzero. Let us postulate that
\begin{equation}\label{eq:U_int}
 U_{\text{int}} = e_N \, \rho \, \xi ,
\end{equation}
which is the simplest possible interaction term. Uncertainties in proportionality factors have been absorbed into the characteristic energy $e_N$ for the interaction per length $b$ between a dislocation line and the nearest grain boundary. We now substitute equation \eqref{eq:U_int} into the expression of configurational free energy density \eqref{eq:F_C}, and take derivatives to find free energy minimum. The result for the steady-state dislocation and grain boundary densities is
\begin{eqnarray}
 \tilde{\rho}^{\text{ss}} &=& \exp \left( - \dfrac{e_D + e_N \tilde{\xi}}{\chi} \right) ; \\
 \tilde{\xi}^{\text{ss}} &=& \exp \left( - \dfrac{e_G + e_N \tilde{\rho}}{\chi} \right) .
\end{eqnarray}
These equations reflect the competition between dislocation formation and recrystallization (or grain boundary formation) in a bid to minimize the free energy. In particular, the first of these equations says that an increase in the grain boundary density $\tilde{\xi}$ reduces the steady-state dislocation density $\tilde{\rho}^{\text{ss}}$. Note that on the right-hand sides of these equations we use the instantaneous values for $\tilde{\rho}$ and $\tilde{\xi}$ instead of their steady-state values.

\subsection{Grain size evolution and rate-hardening anomaly}

Now that we have identified the steady-state grain boundary density or, equivalently, the steady-state grain size, it suffices to write down an evolution equation for the grain boundary density $\xi$ or its dimensionless verion $\tilde{\xi}$. In the linear approximation, $\dot{\tilde{\xi}} \propto (1 - \tilde{\xi} / \tilde{\xi}^{\text{ss}} (\chi))$. The determination of the proportionality factor goes along similar lines as that for the evolution of the dislocation density $\tilde{\rho}$ in equation \eqref{eq:rho}. First, the proportionality factor must be proportional to some rate; the only relevant rate in this problem is the plastic strain rate $\dot{\gamma}_{ij}^{\text{pl}}$. To form an invariant, multiply this by the deviatoric stress $s_{ij}$, which gives the input power associated with shear deformation. Divide this by the reduced shear modulus $\mu_T$ to recover a quantity with the dimensions of inverse time. Thus the evolution equation for $\tilde{\xi}$ reads
\begin{equation}\label{eq:xi}
 \dot{\tilde{\xi}} = \kappa_d \dfrac{s_{ij} \dot{\gamma}_{ij}^{\text{pl}}}{\mu_T} \left( 1 - \dfrac{\tilde{\xi}}{\tilde{\xi}^{\text{ss}}} \right) .
\end{equation} 
The dimensionless quantity $\kappa_d$ specifies the fraction of input power stored in new grain boundaries. It may depend in some complex manner on the other state variables, but for now it may suffice to assume that it is a constant. Converting equation \eqref{eq:xi} to an evolution equation for the characteristic grain size $d = 1 / \xi = b / \tilde{\xi}$, we find
\begin{equation}\label{eq:d}
 \dot{d} = \kappa_d \dfrac{s_{ij} \dot{\gamma}_{ij}^{\text{pl}}}{\mu_T} \dfrac{d}{b} \left( d^{\text{ss}} - d \right).
\end{equation}

The grain size $d$ influences shear-banding dynamics through the parameter $\kappa_1$, which first appeared in equation \eqref{eq:rho}, that specifies the amount of input work stored in newly-created dislocations. It was argued in \cite{langer_2015} that $\kappa_1$ should be a mildly increasing function of the strain rate to describe the rate-hardening anomaly, and that this parameter should be an increasing function of decreasing grain size. \cite{langer_2017b} makes this more explicit by assuming Hall-Petch behavior for this conversion factor. To incorporate both grain size and strain rate dependencies, we propose that
\begin{equation}\label{eq:kappa_1}
 \kappa_1 (d, q) = \kappa_0 + \dfrac{\kappa_r}{\sqrt{d}} \left( 1 + \dfrac{q}{q_r} \right),
\end{equation}
where $q$ is the dimensionless strain rate, introduced above in equation \eqref{eq:q}, $q_r$ is a reference strain rate, and $\kappa_0$, $\kappa_r$ are constants. This closes the feedback between dislocation nucleation, grain size reduction, and strain localization.

\section{Two-zone model for simple shear: results for ultrafine-grained titanium}
\label{sec:4}

\subsection{Simplification to the evolution equations}

In order to tease out the basic physics of adiabiatic shear banding, let us restrict ourselves to a simple two-zone model of a titanium strip undergoing simple shear. The shear rate is $\dot{\gamma} \equiv \partial v_x / \partial y$, and the shear stress is $s_{xy} = s_{yx} \equiv s$, which simplifies the stress evolution equation \eqref{eq:s} considerably:
\begin{equation}\label{eq:s2}
 \dot{s} = \mu (\dot{\gamma} - \dot{\gamma}^{\text{pl}} ) .
\end{equation}
The dimensionless plastic strain rate $q$ in equation \eqref{eq:q} reads
\begin{equation}\label{eq:q2}
 q = \sqrt{\tilde{\rho}} \exp \left[ - \dfrac{1}{\theta} e^{-s / (\mu_T \sqrt{\tilde{\rho}})} \right] .
\end{equation}

The plastic strain rate $q$ in our simple shear geometry varies only as a function of the position $y$ across the material. The two-zone model \cite{langer_2016} provides a further simplification. Use indices 1 and 2 for the shear zone and the rest of the material, respectively, and suppose that the width of the shear zone is a fraction $\alpha$ of the width of the entire strip. The shear stress $s$ is, to a good approximation, uniform across the material; meanwhile, each of the two zones has a uniform plastic strain rate $q_n$, dislocation density $\tilde{\rho}_n$, temperature $\theta_n$, effective temperature $\chi_n$, and characteristic grain size $d_n$, for $n = 1, 2$. The thermal conduction term across the material, described by $\nabla^2 \theta$ above, becomes a simple linear conduction term proportional to the temperature difference between the two zones. Replacing the time variable $t$ by the total strain $\gamma$ for constant loading rate $q_0 \equiv \tau \dot{\gamma}$, the evolution equation for the shear stress becomes
\begin{equation}\label{eq:s3}
 \dfrac{ds}{d \gamma} = \mu \left[ 1 - \alpha \dfrac{q_1}{q_0} - (1 - \alpha) \dfrac{q_2}{q_0} \right] ,
\end{equation}
where
\begin{equation}\label{eq:q3}
 q_n = \sqrt{\tilde{\rho}_n} \exp \left[ - \dfrac{1}{\theta_n} e^{-s / (\mu_T \sqrt{\tilde{\rho}_n})} \right] .
\end{equation}
Equation \eqref{eq:rho} for the dislocation density evolution becomes
\begin{equation}\label{eq:rho3}
 \dfrac{d \tilde{\rho}_n}{d \gamma} = \dfrac{\kappa_1(d_n, q_n)}{\nu_n^2} \dfrac{s q_n}{\mu_T q_0} \left( 1 - \dfrac{\tilde{\rho}_n}{\tilde{\rho}^{\text{ss}} (\tilde{\chi}_n)} \right) ,
\end{equation}
while equation \eqref{eq:chi} for the effective temperature now reads
\begin{equation}\label{eq:chi3}
 \dfrac{d \tilde{\chi}_n}{d \gamma} = \dfrac{\kappa_2}{\mu_T} \dfrac{s q_n}{q_0} \left(1 - \dfrac{\tilde{\chi}_n}{\tilde{\chi}_0} \right) ,
\end{equation}
for $n = 1, 2$. Similarly, the evolution equation, \eqref{eq:d}, for the characteristic grain size becomes
\begin{equation}\label{eq:d3}
 \dfrac{d d_n}{d \gamma} = \kappa_d \dfrac{s q_n}{\mu_T q_0} \dfrac{d_n}{b} \left( d^{\text{ss}}(\tilde{\chi}_n) - d_n \right) .
\end{equation}
Finally, the evolution equations for the temperatures in the two zones now assume the form
\begin{eqnarray}
 \label{eq:theta1} \dfrac{d \theta_1}{d \gamma} &=& K \dfrac{s \, q_1}{q_0} + \dfrac{K_1}{q_0} (\theta_2 - \theta_1) - \dfrac{K_2}{q_0} (\theta_1 - \theta_0) ; \\
 \label{eq:theta2} \dfrac{d \theta_2}{d \gamma} &=& K \dfrac{s \, q_2}{q_0} + \dfrac{K_1}{q_0} (\theta_1 - \theta_2) - \dfrac{K_2}{q_0} (\theta_2 - \theta_0) .
\end{eqnarray}

\subsection{Theoretical analysis and results}

\begin{figure}
\begin{center}
\includegraphics[scale=0.6]{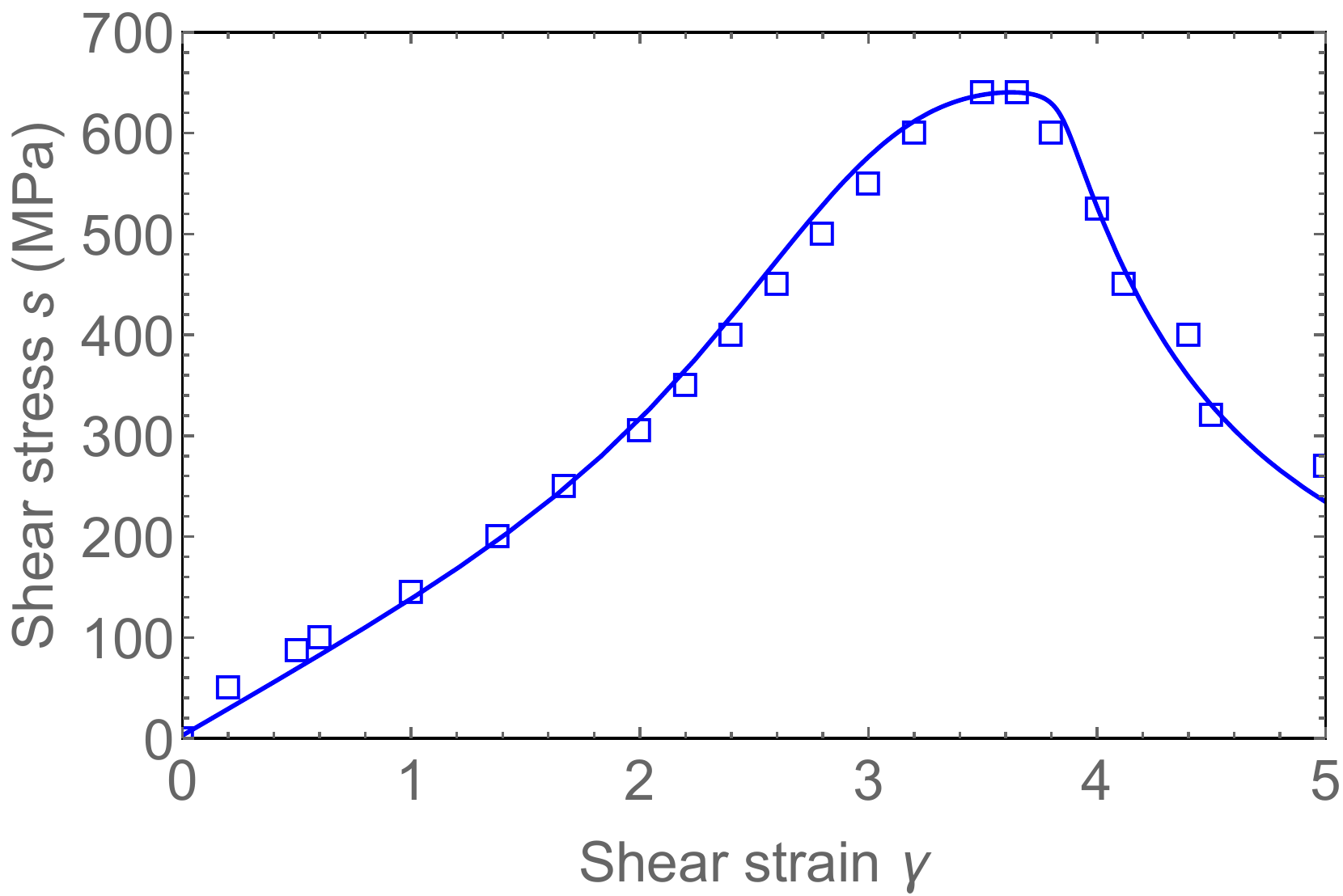}
\caption{\label{fig:splot}Shear stress $s$ as a function of accumulated shear strain $\gamma$, for the titanium strip in question. The applied strain rate is $\dot{\gamma} = 5 \times 10^4$ s$^{-1}$. The solid line represents our calculations, while the squares represent data from experiments on sheared ultrafine-grained titanium \cite{li_2017}.}
\end{center}
\end{figure}

The evolution equations, \eqref{eq:s3} through \eqref{eq:theta2}, are integrated numerically. Figure \ref{fig:splot} shows the shear stress $s$ as a function of shear strain $\gamma$, at a strain rate of $\dot{\gamma} = 5 \times 10^4$ s$^{-1}$. This compares favorably with recent experimental measurements in ultrafine-grained titanium described in \cite{li_2017}, denoted by blue squares in the same figure. In solving these equations, we have used for titanium $\mu = 40$ GPa at temperature $T = 293$ K, and an initial grain size of $d = 120$ nm, taken from \cite{li_2017}. Without loss of generality, the magnitude of the Burgers vector is taken to be the atomic diameter $a$; thus $b = a = 0.14$ nm. To compute the constant $K = \beta / (T_P c_v)$ that describes the amount of work expended in heating up the material, we have used the specific heat per unit mass $c_p = 523$ J kg$^{-1}$ K$^{-1}$, and the density of titanium $\rho_m = 4500$ kg m$^{-3}$, to compute the specific heat per unit volume $c_v$ of titanium, and estimated the Taylor-Quinney factor to be $\beta = 0.9$. Other parameters were estimated somewhat arbitrarily to provide a reasonable fit to the Li et al.~\cite{li_2017} measurement for the shear stress; the parameter values are $T_P = 1.3 \times 10^5$ K, $\mu_T / \mu = 0.06$, $\tilde{\chi}_0 = 0.25$, $e_G / e_D = 0.5$, $e_N / e_D = 50$, $\tau = 10^{-12}$ s, $\kappa_r = 10^{-6}$ m$^{1/2}$, $q_r = 2 \times 10^{-9}$, $\kappa_2 = 1$, and $K_1 = K_2 = 10^{-9}$. There are uncertainties in these numbers, but the approximate relative sizes of many of these parameters are empirically known (e.g.,~\cite{langer_2010,langer_2015}). Our choice of $e_G / e_D = 0.5$ and $e_N / e_D = 50$, for example, imply that the grain boundary energy and the dislocation energy are of the same order of magnitude, and that the interaction energy between grain boundaries and dislocations is only a small fraction of the dislocation energy for typical dislocation and grain boundary densities $\tilde{\rho} \sim \tilde{\xi} \sim 10^{-3}$, which seems to be reasonable. The initial dislocation density, which does not appear to visibly alter the results, is $\tilde{\rho} = 10^{-7}$. To trigger a shear-banding instability, we used the initial effective temperatures $\tilde{\chi}_1 = 0.16$ and $\tilde{\chi}_2 = 0.18$. We also use $\alpha = 0.1$, so that the width of zone 1 is roughly 10 \% that of the strip of interest. As we shall see below, zone 1 is the shear band that accounts for yielding of the material at a shear strain of $\gamma \approx 3.4$. Finally, Figure 3(a) in \cite{li_2017} shows that the stress increases almost linearly with strain, and at times at an increasing rate, for as many as 3 units of shear strain; this is rather unusual in polycrystalline plasticity. To reproduce this behavior, we found it necessary to assume that the hardening parameter $\kappa_0$ in equation \eqref{eq:kappa_1}, which appears in equation \eqref{eq:rho3} and describes the fraction of input work stored in new dislocation lines, depends on the temperature. Here we use $\kappa_0 = \kappa_{00} e^{-\theta_1 / \theta}$ with $\kappa_{00} = 35$ and $\theta_1 = 0.0167$ (equivalently, $\theta_1 T_P = 2000 K$). A temperature-dependent $\kappa_1(d, q)$ may not be the case for other polycrystalline materials (e.g., \cite{langer_2015,langer_2017b}). In order for recrystallization to happen quickly upon yielding, it is necessary to use a temperature-dependent $\kappa_d$ as well. We chose $\kappa_d = \kappa_{d0} e^{- \theta_d / \theta}$ with $\kappa_{d0} = 0.01$ and $\theta_d = 0.0167$. The stress-strain behavior is not sensitive to the choice of $\kappa_d$, though.

\begin{figure}
\begin{center}
\includegraphics[scale=0.6]{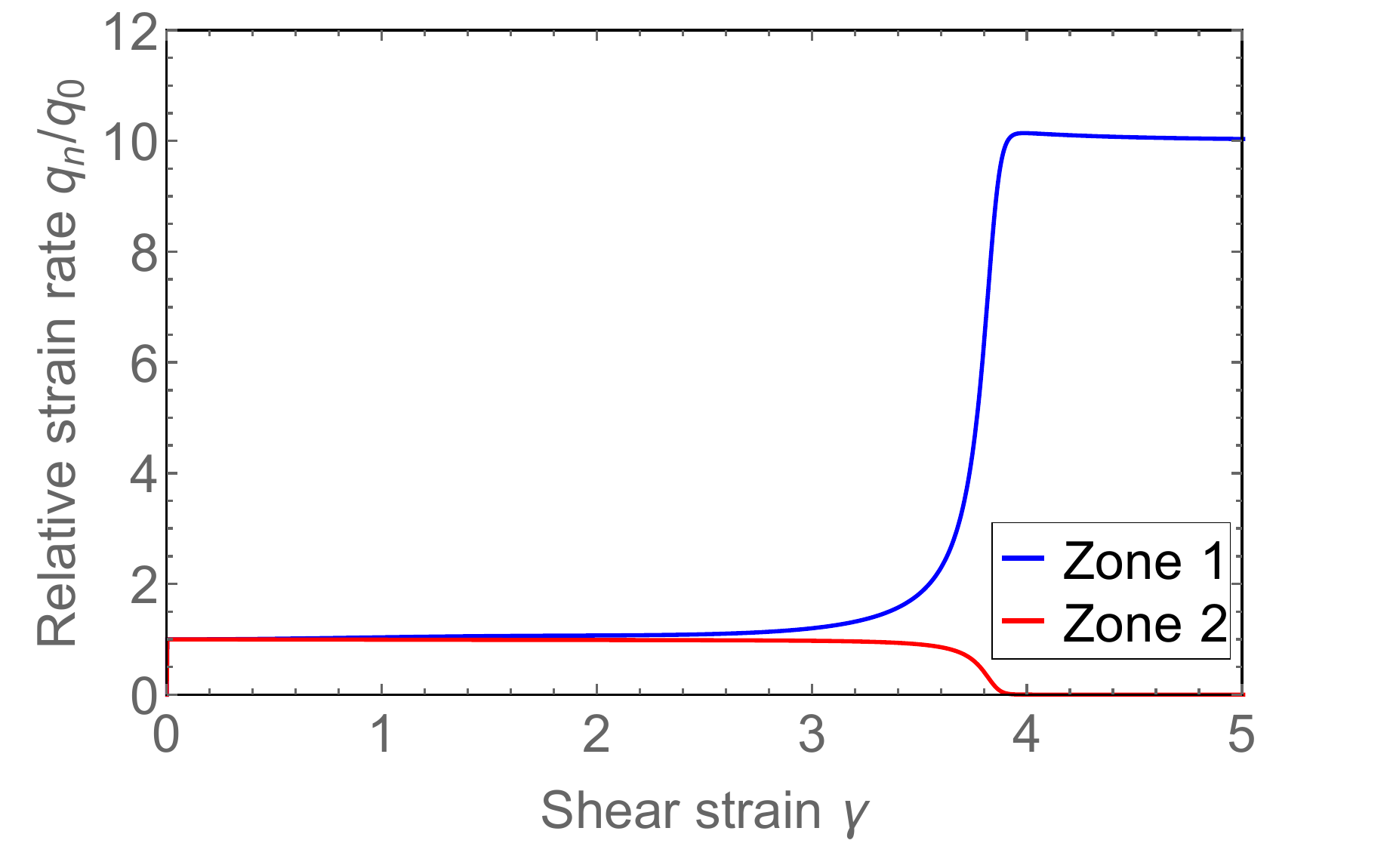}
\caption{\label{fig:qplot}Relative plastic strain rates $q_n / q_0$, for $n = 1, 2$, as a function of accumulated shear strain $\gamma$, for the titanium strip in question. The imposed strain rate is $\dot{\gamma} = 5 \times 10^4$ s$^{-1}$, so that $q_0 = 5 \times 10^{-8}$. It is seen that zone 1 becomes the shear band.}
\end{center}
\end{figure}

Figure \ref{fig:qplot} shows that the strain rate in zone 1, which becomes the adiabatic shear band, stays somewhat higher than that in zone 2 upon the start of plastic deformation; at a shear strain of $\gamma \approx 2.8$, the difference in the plastic strain rates between the two zones starts to grow appreciably, and the shear-banding instability emerges.

\begin{figure}
\begin{center}
\includegraphics[scale=0.6]{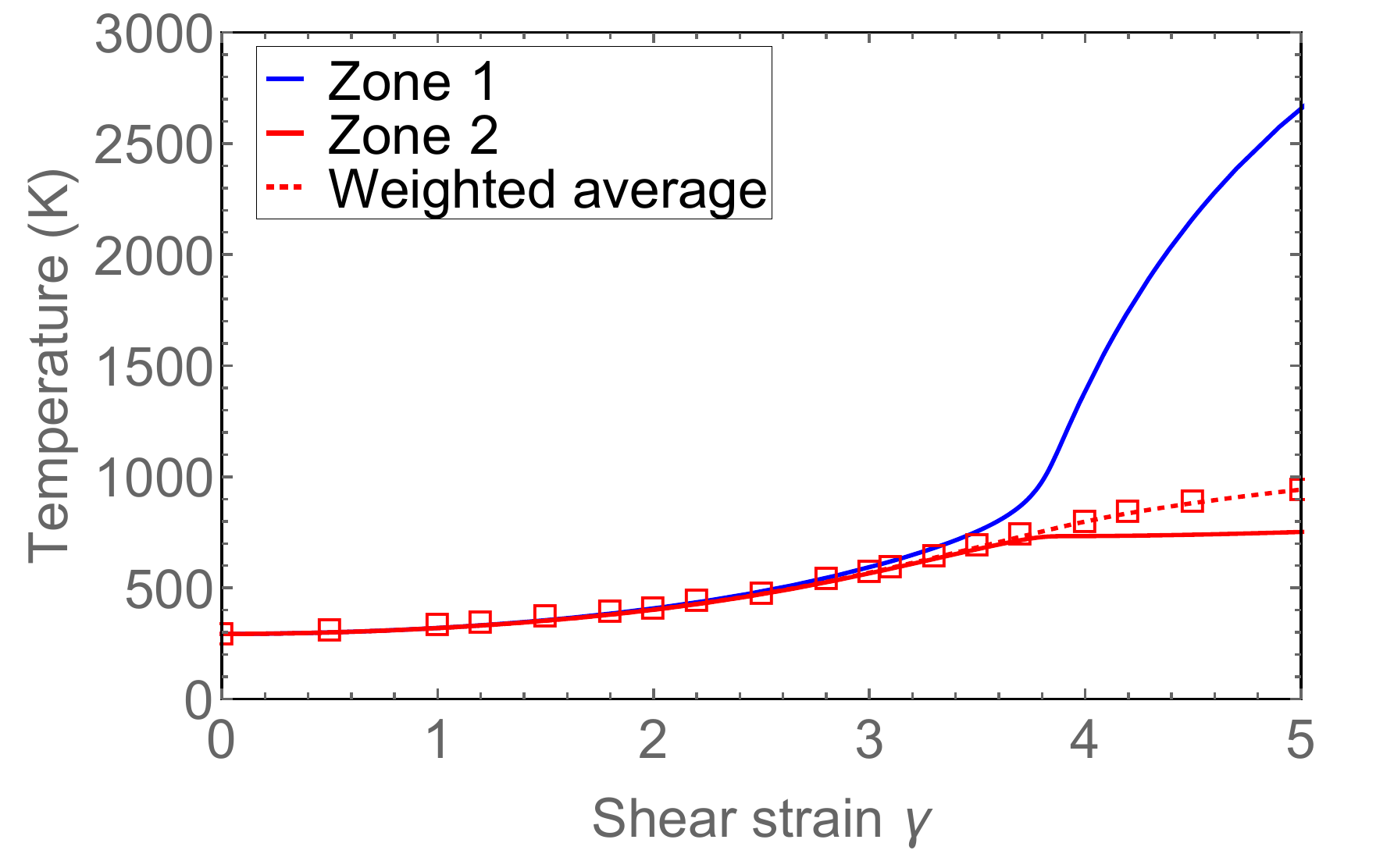}
\caption{\label{fig:Tplot}Temperatures in the two zones $T_n$, for $n = 1, 2$, and their weighted average across the material, as functions of accumulated shear strain $\gamma$, for the titanium strip in question. The applied strain rate is $\dot{\gamma} = 5 \times 10^4$ s$^{-1}$. The lines represent our theoretical calculations, while the red squares represent temperature measurements in the experiments described in \cite{li_2017}. The experimentally-measured temperature lies between the theoretical results for the temperatures within and outside of the shear band, and compares favorably with the weighted average, which is close to the temperature outside the shear band, as it should.}
\end{center}
\end{figure}

To elucidate the origin of the shear-banding instability, turn now to Figure \ref{fig:Tplot}, which shows the temperature rise in each of the two zones in the material. The ``harder'' material in zone 1, which has a lower initial effective temperature, gives rise to a faster increase of the dislocation density and a higher plastic strain rate than that in zone 2. When the strain rate is large enough that the heat generated by the plastic strain cannot diffuse away from zone 1 quickly enough, a shear-banding instability develops. The experimental temperature measurements, shown in the same figure, match the weighted average of the temperatures in the two zones, and is close to the temperature outside the shear band.

\begin{figure}
\begin{center}
\includegraphics[scale=0.6]{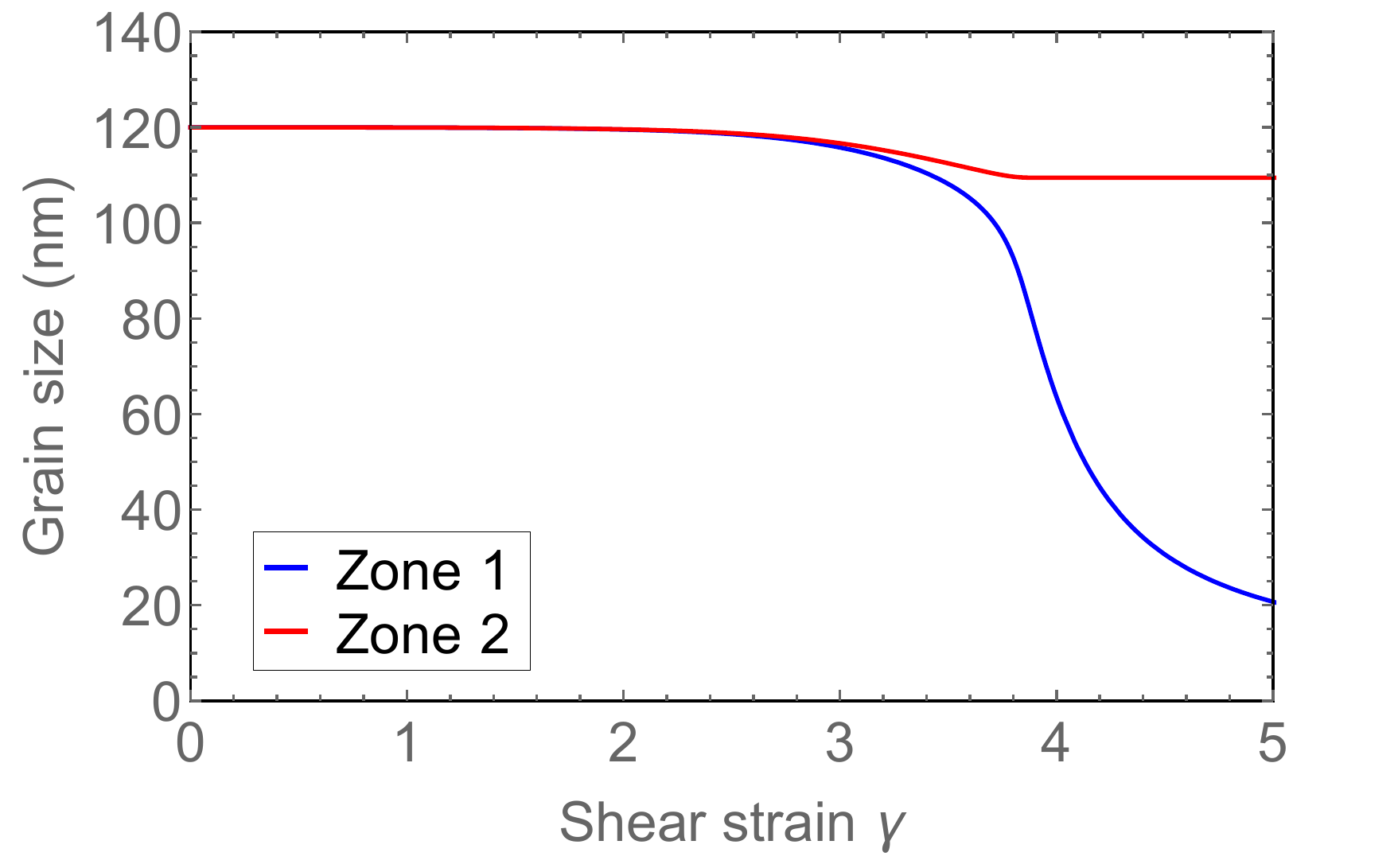}
\caption{\label{fig:dplot}Characteristic grain sizes in the two zones $d_n$, for $n = 1, 2$, as a function of accumulated shear strain $\gamma$, for the titanium strip in question. The applied strain rate is $\dot{\gamma} = 5 \times 10^4$ s$^{-1}$.}
\end{center}
\end{figure}

This is as far as direct comparison with experimental measurements can take us. However, our theoretical analysis, using the same parameters that describe the stress-strain behavior and temperature increase, provides a probe of the structural evolution of the ultrafine-grained titanium, beyond the quantitative measurements reported in \cite{li_2017}. Thus, to study the connection between adiabatic shear banding and dynamic recrystallization, we plot in Figure \ref{fig:dplot} the evolution of the characteristic grain sizes in each of the two zones. The grains in the shear band (zone 1) become much finer upon the onset of shear banding. However, differences in the microstructure evolution between the two zones during the early stages of the deformation is not evident. Moreover, the grain sizes within and outside of the shear band become appreciably different at a shear strain of $\gamma = 3.2$, only shortly before yielding. There is no conclusive evidence, within our calculations, that dynamic recrystallization is a temporal precursor to the shear-banding instability in titanium.

\begin{figure}
\begin{center}
\includegraphics[scale=0.6]{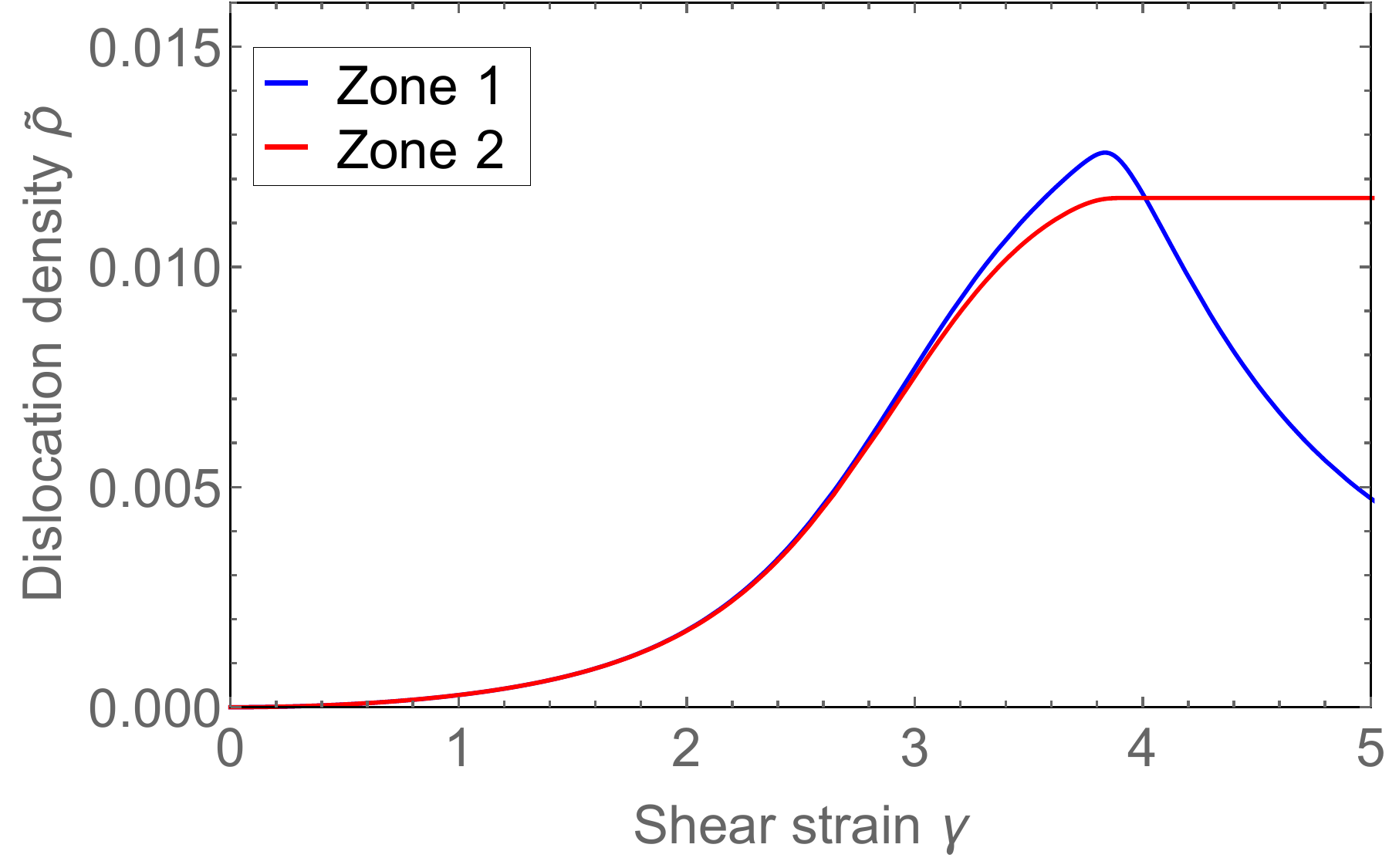}
\caption{\label{fig:rhoplot}Dislocation densities in the two zones $\tilde{\rho}_n$, for $n = 1, 2$, as a function of accumulated shear strain $\gamma$, for the titanium strip in question. Zone 1, the shear band, contains recrystallized grains that are much finer than the grains found in zone 2, as seen in Figure \ref{fig:dplot} above.}
\end{center}
\end{figure}

Figure \ref{fig:rhoplot} compares the dislocation densities within and outside of the shear band. The dislocation density $\tilde{\rho}_1$ within the shear band, where recrystallized grains are subsequently observed, increases somewhat more quickly than outside the shear band. Upon the onset of ASB and recrystallization, however, $\tilde{\rho}_1$ drops markedly, while the dislocation density $\tilde{\rho}_2$ in the rest of the material saturates. Thus our effective-temperature model describes the annihilation of dislocations in the recrystallized grains, or the so-called ``dynamic recovery'', as observed in, e.g., the Rittel et al.~\cite{rittel_2008} experiments. The model shows that this is indeed the entropically favored state.

\subsection{Role of dynamic recrystallization}

\begin{figure}
\begin{center}
\includegraphics[scale=0.5]{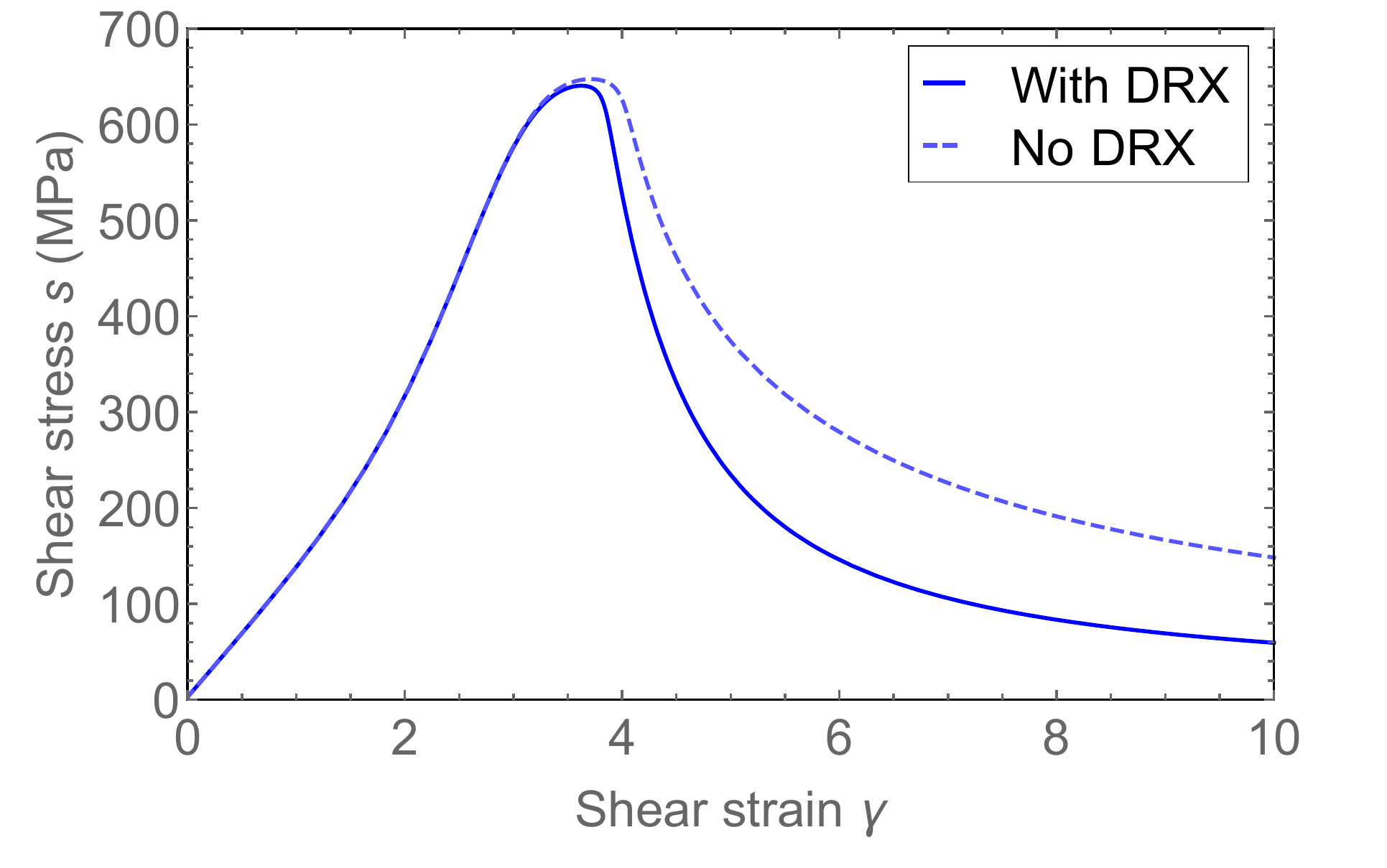}
\caption{\label{fig:splotn}Shear stress $s$ as a function of accumulated shear strain $\gamma$, for titanium which undergoes dynamic recrystallization (solid curve) and pseudo-titanium which does not (dashed curve). The applied strain rate is $\dot{\gamma} = 5 \times 10^4$ s$^{-1}$. Upon the onset of ASB, titanium becomes softer than pseudo-titanium because of DRX.}
\end{center}
\end{figure}

\begin{figure}
\begin{center}
\includegraphics[scale=0.5]{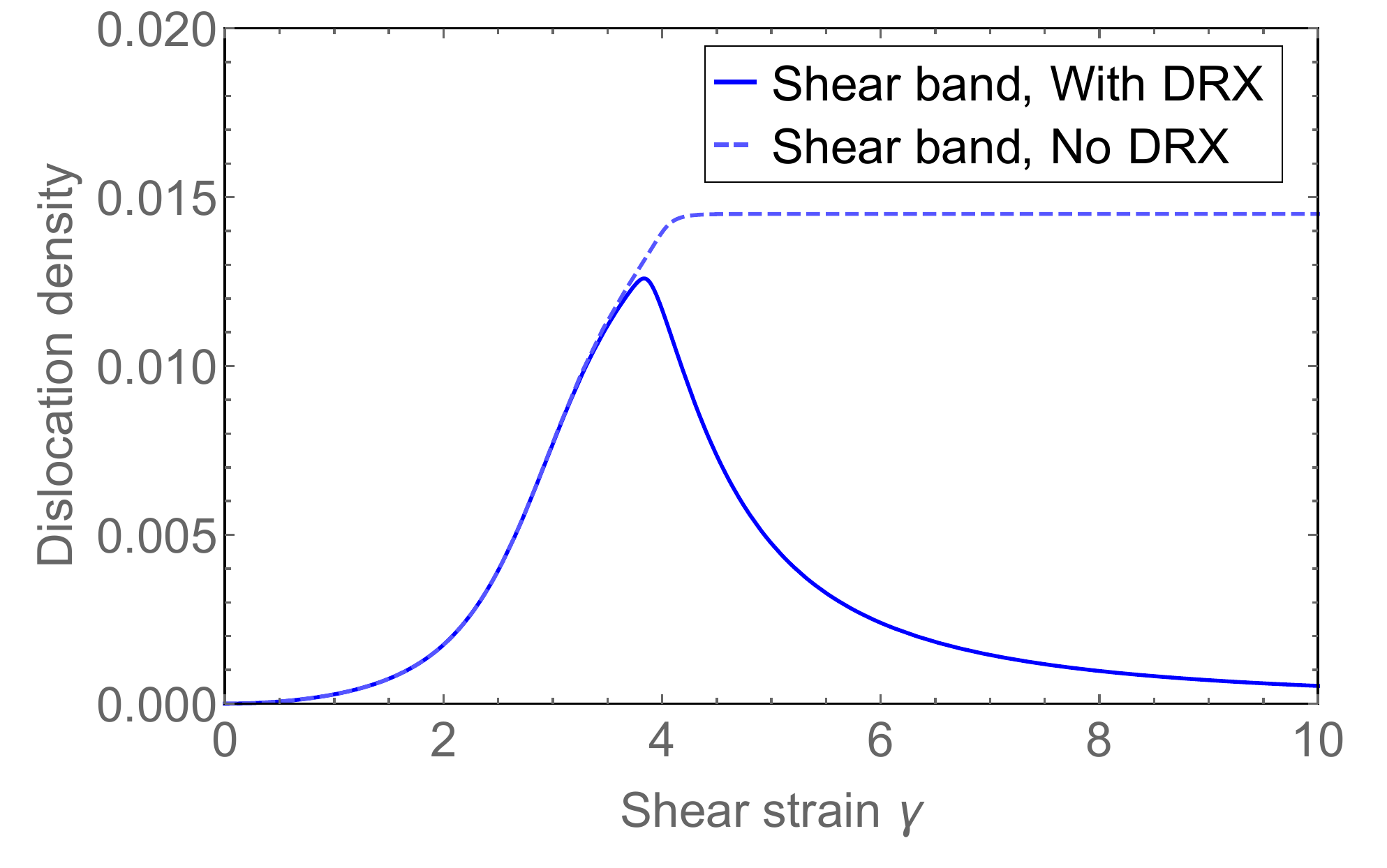}
\caption{\label{fig:rhoplotn}Dislocation densities in the shear band $\tilde{\rho}_1$, as a function of accumulated shear strain $\gamma$, for titanium (solid curve) and pseudo-titanium (dashed curve). The imposed strain rate is $\dot{\gamma} = 5 \times 10^4$ s$^{-1}$. Dislocation annihilation occurs in titanium upon the onset of ASB and DRX. In contrast, the dislocation density saturates in pseudo-titanium, where DRX is prohibited.}
\end{center}
\end{figure}

To elucidate the role of dynamic recrystallization in adiabatic shear banding, we repeat these calculations in an ultrafine-grained ``pseudo-titanium'', whose material parameters are the same as those of ultrafine-grained titanium described above, but where recrystallization is prohibited by setting $\kappa_d = 0$. Figure \ref{fig:splotn} shows the stress-strain curves for titanium and pseudo-titanium at the same shear rate of $\dot{\gamma} = 5 \times 10^4$ s$^{-1}$. It is seen that DRX slightly advances the onset of ASB and failure -- by a shear strain of $\gamma \approx 0.2$ in this case -- and significantly reduces the shear stress upon the onset of ASB when compared to the case in which DRX is prohibited. The advancement of ASB due to DRX can also be seen from a direct plot of the relative plastic strain rates within and outside of the shear band, which we omit here. Thus, DRX is a softening mechanism which, along with thermal heating, prevails over dislocation-induced hardening under suitable conditions, causing material failure through a runaway instability.

Figure \ref{fig:rhoplotn} compares the temporal evolution of the dislocation densities within the shear band for titanium and the DRX-prohibited pseudo-titanium. Whereas the DRX grains in titanium experience dislocation removal, the absence of a DRX mechanism in pseudo-titanium causes the dislocation density there to saturate. This shows that DRX provides a means to dissipate plastic work, and minimizes the free energy more efficiently than dislocation production during the later stages of the deformation, when the dislocation density exceeds some threshold.

\begin{figure}
\begin{center}
\includegraphics[scale=0.5]{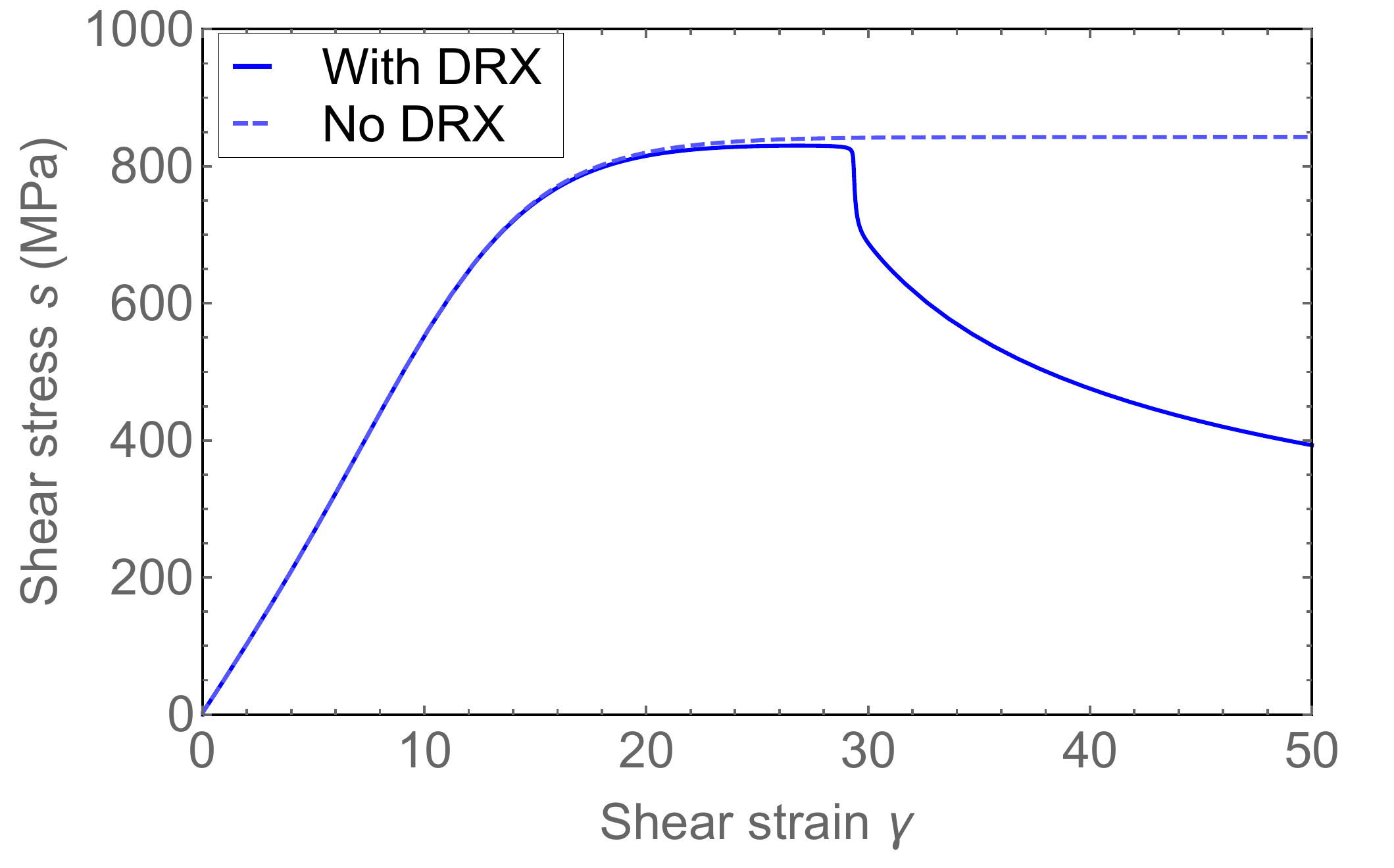}
\caption{\label{fig:splotb}Shear stress $s$ as a function of accumulated shear strain $\gamma$, for titanium which undergoes dynamic recrystallization (solid curve) and pseudo-titanium which does not (dashed curve). The applied strain rate is $\dot{\gamma} = 1.6 \times 10^2$ s$^{-1}$. At this strain rate, ASB develops in titanium but not in the DRX-prohibited pseudo-titanium.}
\end{center}
\end{figure}

It is illuminating to consider slower strain rates. Figure \ref{fig:splotb} shows the stress-strain curves in titanium and pseudo-titanium at the slower strain rate of $\dot{\gamma} = 1.6 \times 10^2$ s$^{-1}$. This is a strain rate at which titanium but not the DRX-prohibited pseudo-titanium exhibits shear banding, as seen in the qualitatively different mechanical behavior. (Even then, ASB does not develop at this strain rate for titanium until the shear strain reaches $\gamma \approx 29$.) This calculation shows that at strain rates slow enough such that the heat produced in the shear band can diffuse away nearly as quickly as it is generated, DRX becomes a necessary ingredient for ASB which would otherwise be suppressed. The role of DRX as a softening mechanism thus becomes evident.

\section{Concluding remarks}
\label{sec:5}

In this paper, we presented an effective-temperature model for microstructural evolution in  polycrystalline materials, and predicted the emergence of shear bands where the material recrystallizes into smaller grains and undergo dynamic recovery. A single, well-defined effective temperature controls the densities $\rho$ and $\xi$ of dislocations and grain boundaries, both being manifestations of configurational, structural disorder. Unlike conventional theories of polycrystalline plasticity, we found no need to make ad-hoc assumptions relating stress, strain and strain rate, or separation into partial stresses for distinct deformation mechanisms. The only necessary assumption, beyond parameter selection, concerns the interaction energy between dislocation lines and grain boundaries. This is however entirely physical because grain boundaries serve as a source of dislocations, and inhibit dislocations from moving between grains. The dislocation and grain boundary densities increase at rates proportional to the plastic work rate, a direct consequence of dimensional analysis and the fact that these structural defects store energy. Shear banding occurs due to pre-existing spatial heterogeneities arising from material preparation. Dynamic recrystallization and dislocation annihilation in the recrystallized grains then emerge naturally in the model, and amount to nothing more than an entropic effect arising from the competition between the formation of dislocations and grain boundaries. We have shown that DRX occurs alongside adiabatic shear banding, and provides a microstructural softening mechanism.

Here, we only considered the dynamics of dislocations and grain size reduction. Other structural changes such as twinning are observed in conjunction with deformation in polycrystalline solids, especially in hcp materials. In the ultrafine-grained titanium measurements reported in \cite{li_2017}, the authors found no signatures of twinning. Moreover, prior experiments in $\alpha$-titanium \cite{salem_2003,salem_2005,salem_2006} show that deformation twinning becomes noticeable in compression but not in shear. Thus we have excluded twinning from the present work. It will be important, however, to investigate how and when to include deformation twinning, which contributes to strain-hardening under compression \cite{salem_2003,salem_2005,salem_2006}, in our present effective-temperature framework.

For simplicity we used a scalar quantity for the grain boundary density $\xi$. Grain elongation occurs in shear bands \cite{bronkhorst_2006}, perhaps prior to recrystallization; \cite{li_2017} proposed a rotational recrystallization mechanism, under which grains rotate in the direction of shear and elongate, before dislocations pileup that creates new grain boundaries for the recrystallized, equiaxed grains. The possible use of a tensorial version of the grain boundary density, which will hopefully address the anisotropic character of this quantity, constitutes future work.

We conclude with a plea for detailed, quantitative microstructural measurements that may help reveal the temporal evolution of the characteristic grain size. This information will help constrain the parameter $\kappa_d$ which quantifies the fraction of input work expended in recrystallization, and shed light the importance of the thermal temperature in DRX. This will, in turn, provide a more rigorous constraint for the Taylor-Quinney factor $\beta$, assumed to be uniformly 0.9 in the present work.

\section*{Acknowledgements}

We thank James Langer and Robert Guyer for instructive discussions. CL was partially supported by the Center for Nonlinear Studies at the Los Alamos National Laboratory over the duration of this work. The authors declare no conflicts of interest. CB was partially supported by the DOE/DOD Joint Munitions Program and and LANL LDRD Program Project 20170033DR.



\bibliographystyle{elsarticle-num} 
\bibliography{actamat_drx_03}





\end{document}